# Interplay of Doping and Structural Modulation in Superconducting $Bi_2Sr_{2-x}La_xCuO_{6+\delta}$ thin films

Z.Z. Li and H. Raffy

Laboratoire de Physique des Solides, Bât. 510, Université Paris Sud, 91405 Orsay, France

S. Bals and G. van Tendeloo

EMAT, University of Antwerp, Groenenborgerlaan 171, Antwerp B-2020, Belgium

S. Megtert

LURE, Bât. 209D, Université Paris Sud, 91898 Orsay, France

Abstract

We have studied the evolution of the structural modulation in epitaxial, *c*-axis oriented, $Bi_2Sr_{2-x}La_xCuO_{6+d}$ thin films when varying the La content $x$ and for a given $x$ as a function of oxygen content. A series of thin films with $0 \leq x \leq 0.8$ have been prepared in-situ by *rf* magnetron sputtering and characterized by R(T) measurements and RBS, TEM and X-Ray diffraction techniques. The oxygen content of each individual film was varied by thermal annealing across the phase diagram. The evolution of the structural modulation has been thoroughly studied by X-Ray diffraction in determining the variation of the amplitude of satellite reflections in special 2 axes $2\theta/\theta$-$\theta$ scans (reciprocal space scans). It is shown that the amplitude of the modulation along the *c*-axis decreases strongly when $x$ increases from 0 to 0.2. It is demonstrated that this variation is essentially governed by La content $x$ and that changing the oxygen content by thermal treatments has a much lower influence, even becoming negligible for $x>0.2$. Such study is important to understand the electronical properties of $Bi_2Sr_{2-x}La_xCuO_{6+\delta}$ thin films.

PACS number(s): 74.72.Hs, 74.78.Bz, 74.62.Dh,

I. INTRODUCTION

Compared with other hole doped superconducting cuprates with one $CuO_2$ plane per unit cell, like $Tl_2Ba_2CuO_6$ (Tl-2201) or $HgBa_2CuO_6$ (Hg-1201), the compound $Bi_2Sr_2CuO_6$ (Bi-2201) exhibits two major differences. The first one is its maximum superconducting critical temperature (20K)[1] [2] [3] which is considerably lower than that of Tl-2201 (80K)[4] or Hg-1201 (94K).[5] For this reason this compound is very important in the investigation of the normal state properties of cuprates and paradoxically to the understanding of the mechanism of high $T_c$ superconductivty. The other difference, probably correlated with the first one, is the fact that this compound exhibits an incommensurate structure modulation in the b-c plane of displacive behaviour.[6] To understand why the Bi-2201 system is so different from other cuprates, the majority of the published works consider the effect of doping[7] [8] [9] which can be achieved in two ways. The first one is by changing the oxygen content in the system.[3] This is sometimes called *mobile* doping.[10] The second way is by using cationic substitution, such as substituting divalent Sr by trivalent La which decreases the hole content, or trivalent Bi by divalent Pb which produces the reverse effect.[2] [11] This can be called *non mobile* doping. Doping studies show that Bi-2201 is one of the few systems which can be completely investigated from strongly underdoped to metallic overdoped states.[3] [12] The $T_c$ value of pure Bi-2201 can vary from 0 to a maximum value $T_{cmax}$ at most equal to 20K, and which depends strongly on the Bi/Sr ratio. The value of $T_c$ can be raised by non mobile doping up to 30-35K using La/Sr substitution and even up to 45K with simultaneous La/Sr and Pb/Bi substitution.[13] [14] The origin of the structural modulation present in the Bi-2201 system is often thought to be due to the mismatch between the BiO layers and the perovskite slab and/or the presence of extra oxygen.[15] However the relation between the modulation and the extra oxygen is not clear, even controversial. On one hand some authors[16] suggest that there is no essential relationship between the incommensurate modulation and extra oxygen. On the other hand, other authors[17,18] suggest that the extra oxygen is the real driving force behind the modulation. In fact, the samples studied were only substituted with La in the former case and only oxygen doped in the latter. Therefore it seems clear that in order to find a possible relation between the modulation and extra oxygen, it is necessary to systematically study both doping situations.

In this paper, we report our results obtained in a systematic study of both non mobile La doping (from x = 0 to x = 0.8) and mobile doping for each La content in Bi(La)-2201 thin films. The use of thin films allows us to easily modify the oxygen content of the samples and to perform transport measurements. The mobile doping levels of these thin films were varied from overdoped to underdoped. The main goal of this study is to show the relation between extra oxygen and the modulated structure for each given La content in the Bi-2201 thin films. The most important issue in the present study is that we show how much La doping influences the amplitude of the structural modulation in these thin films whatever the oxygen doping.

II. EXPERIMENTAL

Epitaxial *c*-axis oriented $Bi_2Sr_{2-x}La_xCuO_{6+y}$ ($x$ = 0 to 0.8) thin films, 2000 to 3000 Å thick, were prepared by single target reactive *rf* magnetron sputtering on heated (720-750°C) single crystal $SrTiO_3$(100) substrates and also on MgO(100) substrates for composition analysis. The deposition conditions were similar to those previously described for pure Bi-2201 thin films.[3] To prepare films with different *x* values, a series of targets with the required *x* values were used. The composition of the films measured by RBS on films deposited on MgO is shown in table I. The measured values of *x* are quite close to the nominal composition of the target except for *x* = 0.8 ( about 15% more in the films). Fig. 1 presents a [110] HRTEM image of such a film viewed in cross section. The quality of the film/subtrate is perfect. There is complete epitaxy, no amorphous layer between film and substrate and no secunary phases formed at the interface.

After deposition the samples were cooled down to 420°C under process gas. For each *x* value, the films were studied in four typical oxygen doping states, spanning the phase diagram, obtained as follows. The so-called *as-prepared* state was obtained in-situ: it is close to optimal doping for samples with $0 \leq x \leq 0.4$, which were kept in vacuum at 420°C for one hour, while it is an underdoped state for the samples with $0.5 \leq x \leq 0.8$ although they were continuously cooled down under $O_2$ rich process gas. The three other doping states were subsequently obtained by ex-situ annealing treatments of the same sample in different oxygen atmospheres: i) The samples were partially deoxygenated by an annealing treatment in vacuum at 300°C for one hour giving underdoped states. ii) Then the samples were oxygenated in a pure oxygen flow at 420°C. iii) Finally the samples were exposed in an oxygen plasma at 420°C with 100% $O_2$ to obtain a still higher oxygenation level. After each treatment, the samples were characterized by the following methods. The superconducting transition temperatures of the films were measured resistively by a standard *dc* four-probe method and also confirmed magnetically by *ac* susceptibility measurement with a SQUID magnetometer. The properties of the crystal structure of the films were thoroughly studied by X-ray diffraction (XRD): conventional *θ-2θ* scan, rocking curve scan, partial pole figure (*ϕ*-scan) and special 2 axes *2θ/θ-θ* scan (reciprocal scan) methods were used.

To study the structural modulation in the Bi 2201 system on a local scale, transmission electron microscopy (TEM) has proved to be very useful[11,19]. TEM is certainly a very powerful method, but as it is a destructive technique it does not allow the examination of the evolution of the modulation in the same sample for different oxygen doping levels. In the present case (a Bi-2201 film on a STO substrate) the preparation of suitable cross section TEM samples was not straightforward because of the limited adhesion of the film to the substrate. An alternative approach is to use X-ray diffraction techniques (XRD). For ceramic samples (powder), the structural modulations were studied from standard $\theta$-$2\theta$ diffractograms.[20] However this technique is not suitable for thin films because it appears that the modulation amplitude is much less important there than in ceramic samples resulting in very low diffracted intensities of the same order of magnitude as the background in between average structure Bragg peaks that are difficult to measure. The situation is worse when the films are highly textured or even epitaxied. For single crystal samples, the modulated structures were studied using a precession camera technique.[7] This method is also difficult to use for thin films due to the presence of thick absorbing $SrTiO_3$ substrate. The diffraction volume ratio between film and substrate is inadequate for this transmission technique. We then used reflection diffraction geometries on a 4-circle diffractometer. Copper radiation was used for this study. A graphite analyser was mounted in front of the detector (proportional gas counter). As XRD is a non destructive technique (at least for the kind of samples studied here), we were able to examine the modulated structure of the same sample at different oxygen doping levels.

III. RESULTS AND DISCUSSION

Figures 2a and 2b show the temperature dependence of the resistance of $Bi_2Sr_{2-x}La_xCuO_y$ films with $0 \leq x \leq 0.6$ in the as-prepared conditions. Critical temperature values, $T_c(R=0)$, are about 20K for $x = 0, 0.05, 0.1$, and $T_c(R=0)$ values are up to 30K for $x = 0.2, 0.3, 0.4$.[21] On the other hand $T_c(R=0)$ is equal to 20K for $x = 0.5$ and is less than 4.2K for $x = 0.6$ and for $x = 0.8$, the R(T) curve (not shown) exhibits a semi-conducting-like behaviour. It is thought that maximum $T_c$ in the La doped Bi-2201 system occurs for $x$ equal to about 0.4 [22] Therefore the region where $x$ is less than 0.4 is called overdoped. Our results presented in the insert of Figure 2b show that the maximum $T_c$ corresponds to $x$ = 0.2, 0.3 and 0.4, that is $T_c$ is maximal over a large compositional range. It is known that in this system $T_c$ is not only dependent on $x$, but is also dependent on the Bi/Sr ratio. The microstructure plays an important role in the value of $T_c$ as well. In the rest of this paper, we will see that the microstructure distortion due to the incommensurate modulation is attenuated by La doping for $x \geq 0.2$. So the large modulation amplitude existing for $x \leq 0.1$ could be one of the main reason to explain why $T_{cmax}$ is only 20K in these cases. It should be noted that a phase transition from monoclinic to orthorhombic occurs for $x$ equal to about 0.1[7]. Figure 3 allows us to compare the $T_c$ values of the as-prepared (or in situ) samples with the values found after the three annealing treatments: deoxygenation in vacuum, oxygenation in $O_2$ flow and in oxygen plasma. It appears that for $0 \leq x \leq 0.4$ the as-prepared samples were optimally doped as $T_c$ decreases under ex-situ annealing. This is in agreement with the linear variation of $R(T)$ in a large $T$-interval (Fig.2a). In contrast, for $x = 0.5$, $T_c(R=0)$ appears not to be influenced by oxygen overdoping. The reason for this remains unclear. It should be noted that in Figure 4 of reference 22 the $T_c$ value for the same composition is out of the parabola line representing $T_c(p)$, and in reference[23], $T_c$ for $x$ equal to 0.5 was only 10K whereas it was 33K for $x = 0.4$. For $x = 0.6$, the samples were underdoped as $T_c(R=0)$ value increased from less than 4.2K (as-prepared sample) to 8K and to 12K when the samples were subsequently treated by annealing in oxygen atmosphere and by exposing in pure oxygen plasma respectively. For $x=0.8$, after the oxidising treatments, the film remains semi-conducting at low temperature although there is a decrease of the resistance values.

Phase identification, crystal orientation and film structure were obtained from XRD $\theta$-$2\theta$ scans. In all cases the XRD patterns of the films correspond to a single phase as can be seen on a typical XRD pattern given in Figure 4. The full width at half maximum (FWHM) of the rocking curves through the (*008*) diffraction peak were always in the range of 0.15 to 0.2° indicating that all films are highly *c*-axis oriented (with *c*-axis perpendicular to the film and substrate surfaces). The epitaxial character is further attested to with the recording of a partial pole phi-scan of the (*115*) Bragg reflection as shown in Figure 4. Knowing the parameter *c* from the earlier *θ-2θ* scan, the lattice parameter *a* is deduced from another *θ-2θ* scan carried out in the reciprocal direction (*h* 0 *3h*). The variations of lattice parameters *c* and *a* of the films as a function of La content are given in Figures 6a and 6b. It is shown that *c* decreases and *a* increases with increasing La content. From Figure 6a, we can also see that *c* decreases with increasing oxygen doping level for given La content (see also Fig.2 of Ref 12). This result is in agreement with those reported for single crystals and for ceramic samples.[7][2] In general, the decrease of the lattice parameter *c* is thought to derive from the fact that the atomic radius of La is smaller than that of Sr. But from our results or from those of references 7 or 2, the variation of the parameter *c* between *x* = 0 and *x* = 0.5 for instance is much larger than the simple difference of atomic radius between La and Sr. So this explanation is not sufficient. In fact, as we know, in the Bi-2201 system, the lattice parameter *c* decreases while increasing the $O_2$ doping level. The reason for this could be the insertion of extra oxygen in the BiO bilayer. If this is correct, the decrease of lattice parameter *c* in the La-doped Bi 2201 system is of the same origin: additional oxygen atoms needed for charge equilibrium increase with increasing La content and are also inserted in the BiO layers.

Let us now consider the main results of this X-Ray diffraction study. In order to investigate the incommensurate structure modulation in the La-doped Bi-2201 system, scans in reciprocal space were carried out around the (*0 0 16*) main Bragg reflection. Typical patterns of this kind of scan are shown in Figures 7a, b, c. These figures show the presence of modulation in all of the Bi(La)-2201 system as signalled by the existence of satellite reflections around a main Bragg peak of the average structure. The position and shape of the observed satellites are related to the La content of the as prepared films. In the reciprocal space, the modulation wave vector can be described as $q = (0, q_b, q_c)$. Figure 7c shows that for $x \geq 0.3$, the shape of the satellite contour becomes elongated. This fact means that there is a loss of long range order of the atomic displacements along the c-axis. The figures 8a and 8b show the variation of $q_b$ and $q_c$ as a function of La and $O_2$ doping level. In the case of La-undoped Bi-2201 as-prepared samples only, there is a second kind of modulation with q' = (0, 0.17, 1), that disappears once the samples are annealed (this modulation is no longer present for La-doped even with $x$ as low as $x = 0.05$). We can see that $q_b$ and $q_c$ increase monotonically with increasing La content. The results about $q_b$ are very similar to those of N. L. Wang et al.[24] For instance, the values of $(q_b(x) - q_b(x = 0)) / q_b(x = 0)$ were about 13, 18 and 20% for $x = 0.2$, 0.3 and 0.6 respectively in their case and the values were about 15, 17 and 19% respectively in our case. However the superconducting properties are quite different between their ceramic samples and our thin films. We can see also from Fig.8 that for $x = 0$ and 0.05, the values of $q_b$ and $q_c$ increase with increasing $O_2$ doping level. In contrast, for $x \geq 0.1$, $q_b$ and $q_c$ becomes independent on the $O_2$ doping level.

Two samples (x=0 and x=0.3) have been examined by TEM (fig. 9). Measurement of the q-vectors by electron diffraction reveals q = (0, 0.18, 0.34) for the x = 0 compound and q = (0, 0.22, 0.82) for x = 0.3. With an accuracy of ±0.03, these values fit very well with the values determined by X-ray diffraction (see fig. 8).

As is well known in the case of displacive modulated structures, satellite intensities are related to the amplitudes of the atomic displacements of those atoms that are involved. For small displacements relative to the cell parameters, the contribution of one kind of atom to the amplitude of the diffracted wave is proportional to the scalar product of its displacement vector R = ($u, v, w$) with the scattering vector Q = ($h, k, l$). The total diffracted amplitude can be complicated to calculate because of the summation over all of those atoms that are involved in the modulated structure. But what is left is that the corresponding diffracted intensity (the square of the modulus of the diffracted amplitude) will be low if the displacements are very small, higher if the displacements are larger, but always proportional to the square of the scattering vector length. Interestingly, as we measured satellites intensities around the Bragg reflection (*0 0 16*), we got insight into those atomic displacements that are polarized along *c* direction, that is perpendicular to the superconducting $CuO_2$ planes. Therefore, to measure the behaviour of the amplitude of the modulation at large, with respect to the different doping conditions, we took the following quantity as a test probe:

$$(I_s - I_b) / (I_m - I_b).$$

where $I_s$ and $I_b$ are the average values of the integrated intensities of the satellites and the neighbouring background respectively and $I_m$ the related main Bragg peak intensity. Figure 10 shows the variation of the modulation amplitude as a function of La content and doping level. The results demonstrate that the modulation amplitude decreases with increasing La content up to *x* = 0.3. Then above *x* = 0.4 the amplitude increases again slightly. It should be noted that the value of the lattice parameter *a* for the smallest modulation amplitude coincides with that of Bi-2212 compound, where the amplitude of the modulation is known to be very small in thin films. It is also to be noted that for a given *x* the modulation amplitude increases with increasing $O_2$ doping level for *x* = 0, 0.05 and 0.1. In contrast for *x* ≥ 0.2, the amplitude appears to be independent of the $O_2$ doping level. Whatever the oxygen content the minimum of the modulation amplitude occurs for x=0.3-0.4. So from the comparison of Fig.10 and Fig. 3 giving $T_c(x)$, there appears to be some correlation between the occurrence of the maximum of $T_c(x)$ and the minimum of the modulation amplitude.

Several models aiming at the explanation of the origin of the structural modulation are summarized in reference 15. Here we focus only the relation between the extra oxygen and the modulation. From Fig.8 and Fig.10, it can be seen that it is only for x<0.1 and for the higher oxygenation levels that the modulation depends on the extra oxygen. In the La doped Bi-2201 compounds, the extra oxygens located in the double BiO planes can be divided into two kinds of oxygen. One is used for doping the $CuO_2$ planes and the second one, so called additional oxygen, is used for charge equilibrium as a consequence of the substitution of the divalent Sr by trivalent La atoms. For x<0.1, the extra oxygen is mostly used for doping the $CuO_2$ planes. The higher oxygen doping level states with no superconducting transition down to 4.2K can be easily obtained. However, further increase of the La content leads to increase the additional oxygen, which occupies more and more the BiO planes[8]. This may be the reason why the values of $q_b$ and $q_c$ increase with increasing La content (Fig. 8). The fact that additional oxygen occupies more and more room leads to the fact that the space left in the $Bi_2O_2$ bilayer for the oxygen used for doping the $CuO_2$ planes is less and less important. Consequently oxygen overdoping become more and more difficult. As a matter of fact, for 0.1<x<0.5 the most oxygenated states always exhibit a superconducting transition and their $T_c$ values increase with x (Fig. 3), and for x>0.6 the overdoped region even cannot by reached by oxygenation.

In summary, by a systematic study of both non mobile La doping (from 0 to 0.8) and mobile $O_2$ doping for each given La content, in Bi-2201 thin films, our main results show that: 1) $T_c$ optimum corresponds to $0.2 \leq x \leq 0.4$ and not only to $x = 0.4$. 2) For x<0.1 and for the higher oxygen doping levels the modulation varies with varying oxygen doping level. In this case, we agree with the authors of references 17 and 18, upon the fact that the extra oxygen is the real driving force of the modulation. For x>0.1, the modulation becomes independent of the $O_2$ doping level. Then we agree with reference 16 upon the fact that there is no essential relation between the incommensurate modulation and doping. Last but not least, we have shown that the amplitude of the modulation along the c-axis decreases while La content increases from 0 to 0.3. This result is very important to understand the electronic coupling between $CuO_2$ layers of the Bi(La)-2201 system and the related anisotropy .[25]

Acknowledgments.
We would like to thank N. Blanchard for powder synthesis, D. Petermann for software support in X-Ray analysis and F. Lalu, CSNSM, for RBS studies.

Table 1

**Nominal composition of the Bi(La)-2201 targets with various La content and composition of the films measured by RBS**

| La | Target (Nominal) | Films (RBS) |
|---|---|---|
| x | Bi: Sr: La: Cu | Bi: Sr: La: Cu |
| 0 | 2.00:1.90:0.00:0.90 | 2.07:1.93:0.00:1.20 |
| 0.05 | 2.00:1.95:0.05:0.90 | 2.20:1.93:0.07:1.04 |
| 0.1 | 2.00:1.90:0.10:0.90 | 2.10:1.90:0.10:1.14 |
| 0.2 | 2.00:1.80:0.20:0.90 | 2.00:1.82:0.18:1.10 |
| 0.3 | 2.00:1.70:0.30:0.90 | 1.94:1.72:0.28:1.08 |
| 0.4 | 2.00:1.60:0.40:0.80 | 1.91:165:0.35:1.21 |
| 0.5 | 2.00:1.50:0.50:0.80 | 2.00:1.50:0.50:1.01 |
| 0.6 | 2.00:1.40:0.60:0.90 | 1.79:1.40:0.60:1.40 |
| 0.8 | 2.00:1.20:0.80:0.90 | 1.82:1.14:0.90:1.22 |

**Fig.1: [110] HRTEM image of an undoped Bi-2201 thin film. The cations are imaged as bright dots under the present conditions.**

**Fig.2a and Fig.2b**: Temperature dependence of the resistance normalized to its value at 300K, $R/R_{300K}$, for x=0 to 0.6, of $Bi_2Sr_{2-x}La_xCuO_y$ thin films in the as-prepared state. In insert of Fig. 1 a, $T_c(R=0)$ as a function of $x$ for $Bi_2Sr_{2-x}La_xCuO_y$ thin films in the as-prepared state.

**Fig.3**: $T_c(R=0)$ vs. $x$ of Bi(La)-2201 films in the different doping states : ○: as prepared. □: Annealed at 300°C in vacuum. ♦: Annealed at 420°C in oxygen flow. ●: Exposed at 420°C in pure oxygen plasma.

**Fig.4**: A typical XRD pattern for a Bi(La)-2201 film. The reflection peaks of $SrTiO_3$ substrate are indicated (the square root of the intensity is plotted to enhance the visibility of low intensity diffraction peaks).

**Fig.5**: XRD φ-scan of the (*115*) reflection of a Bi(La)-2201film.

**Fig.6a and Fig.6b**: Variation of the lattice parameters *a* and *c* as a function of $x$ for $Bi_2Sr_{2-x}La_xCuO_y$ thin films in the different doping states (see text).

**Fig.7a, 7b, 7c**: XRD $\theta/2\theta-\theta$ scan patterns of Bi(La)-2201 films with $x$=0, 0.2, 0.3 respectively, in the as-prepared state.

**Fig.8a, 8b:** $q_b$ and $q_c$ v.s. $x$ of Bi(La)-2201 films in the different doping states (see text).

**Fig.9:** [110] electron diffraction pattern of a) the undoped Bi-2201 and b) the Bi(La)-2201 film with x = 0.3.

**Fig.10:** Reduced satellite intensity $(I_s - I_b) / (I_m - I_b)$ vs. $x$ of Bi(La)-2201 film in the different doping states. Subscript -s- is for satellite, -m- for main Bragg reflection peak, -b- for the nearby background.

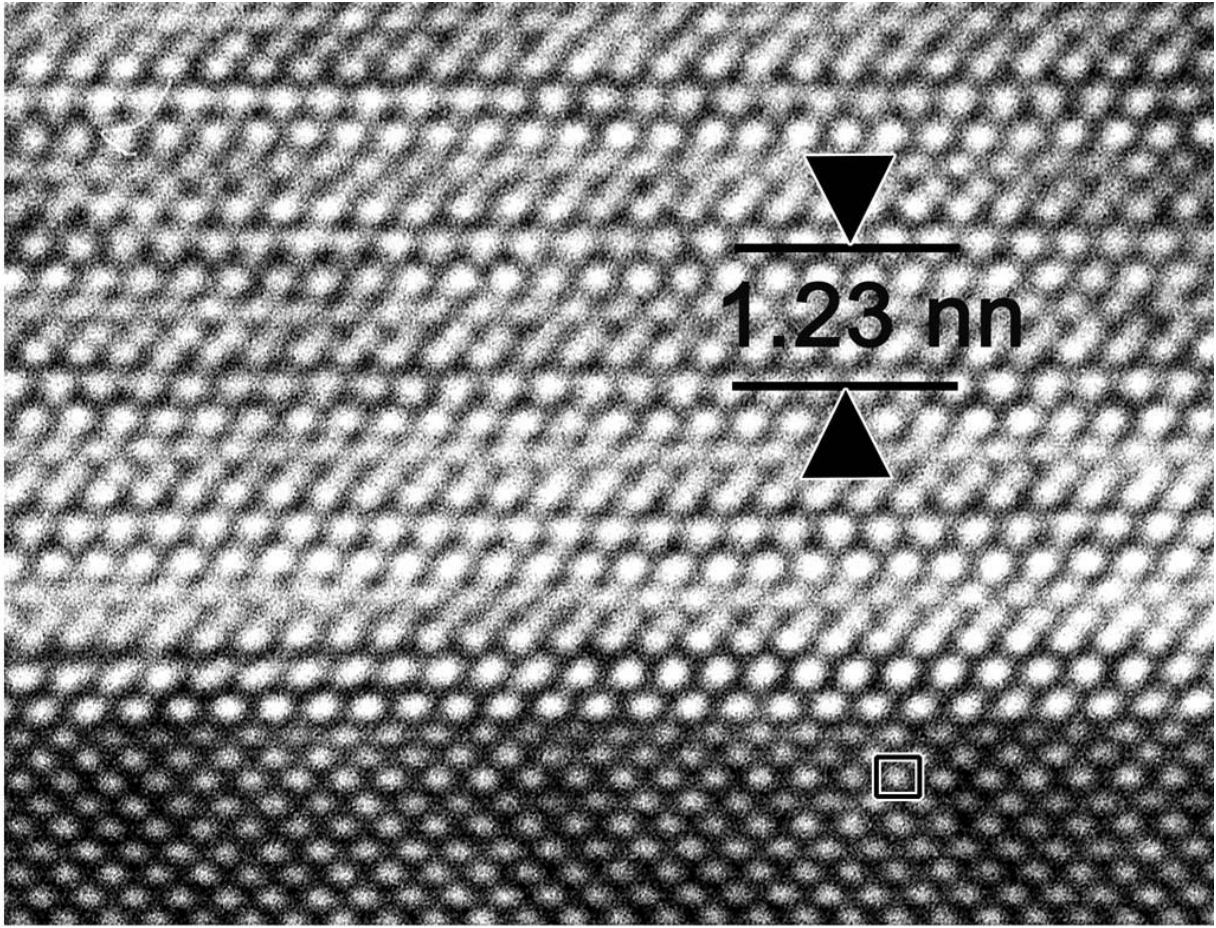

Z.Z.Li *et al.*Fig.1

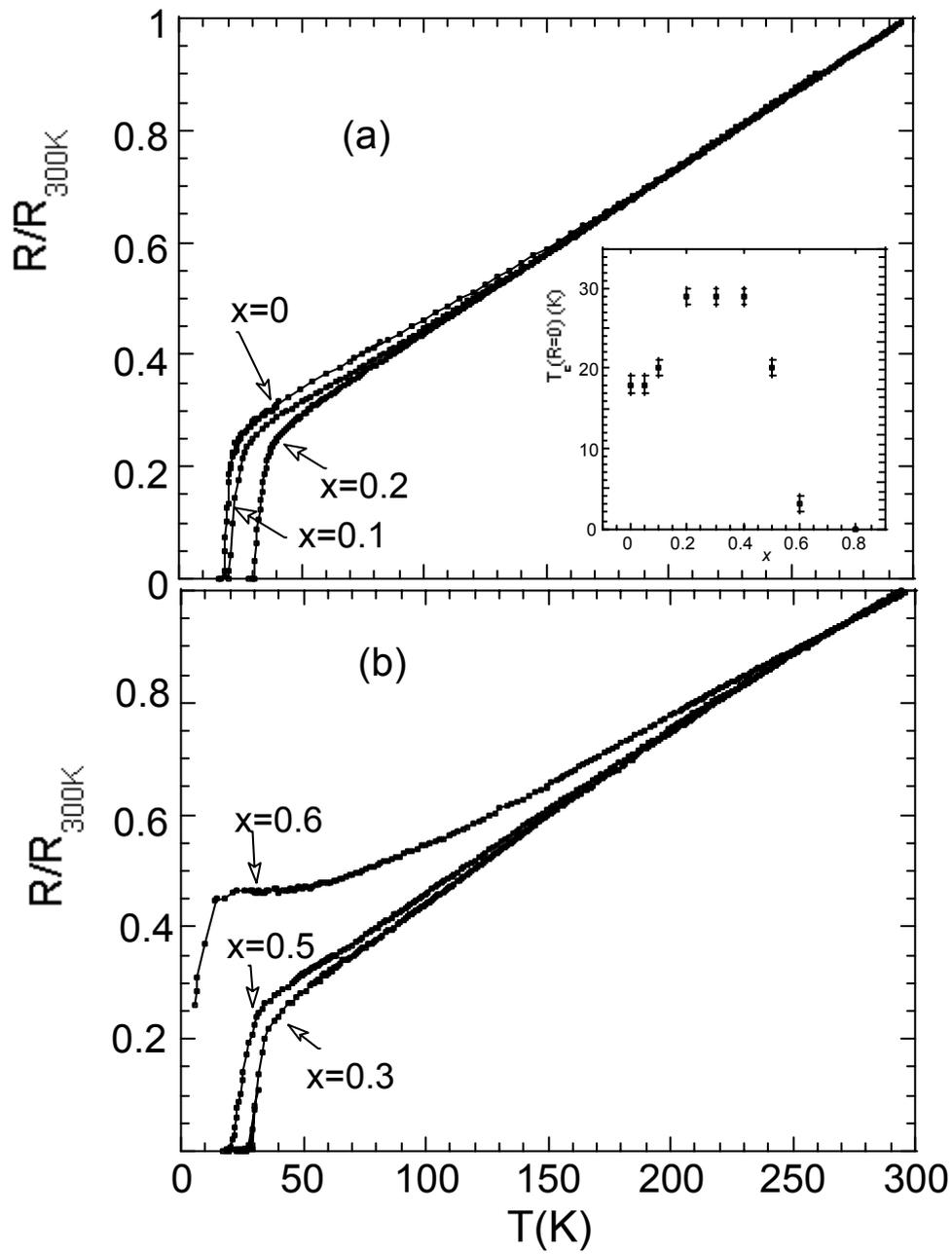

Z.Z. Li *et al*. Fig.2

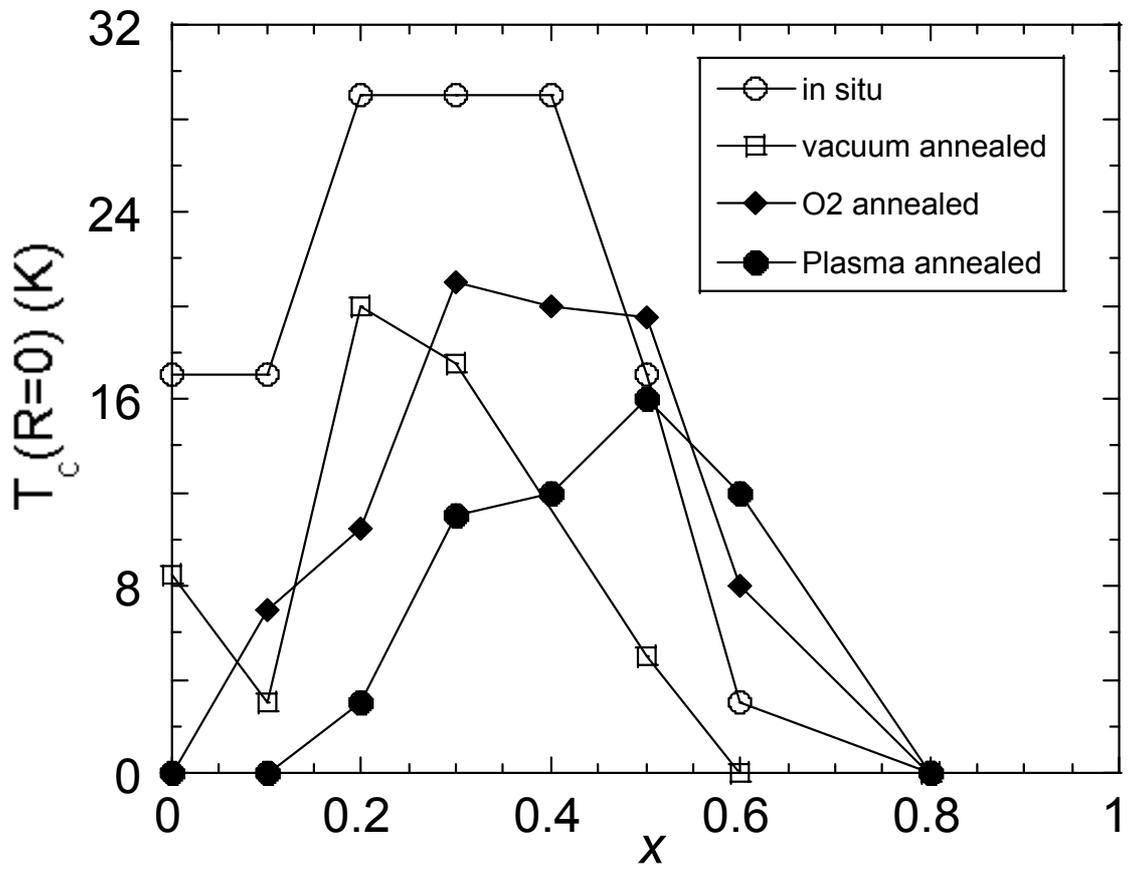

Z.Z. Li *et al.* Fig.3

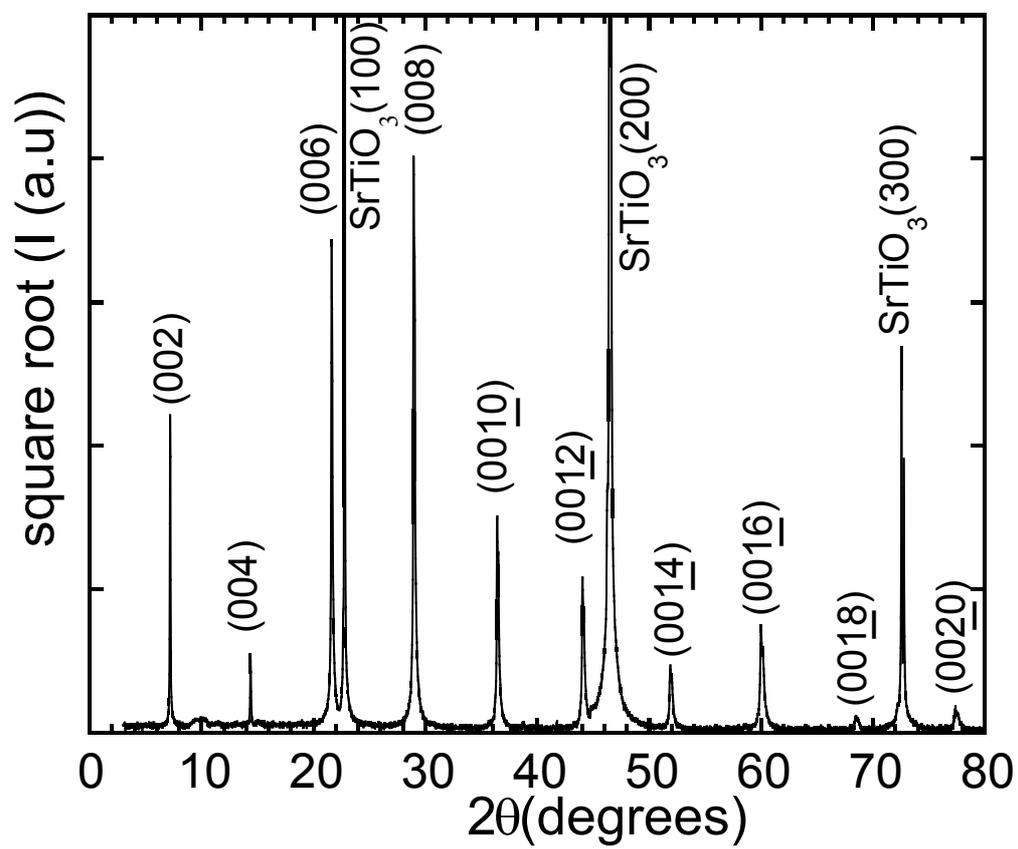

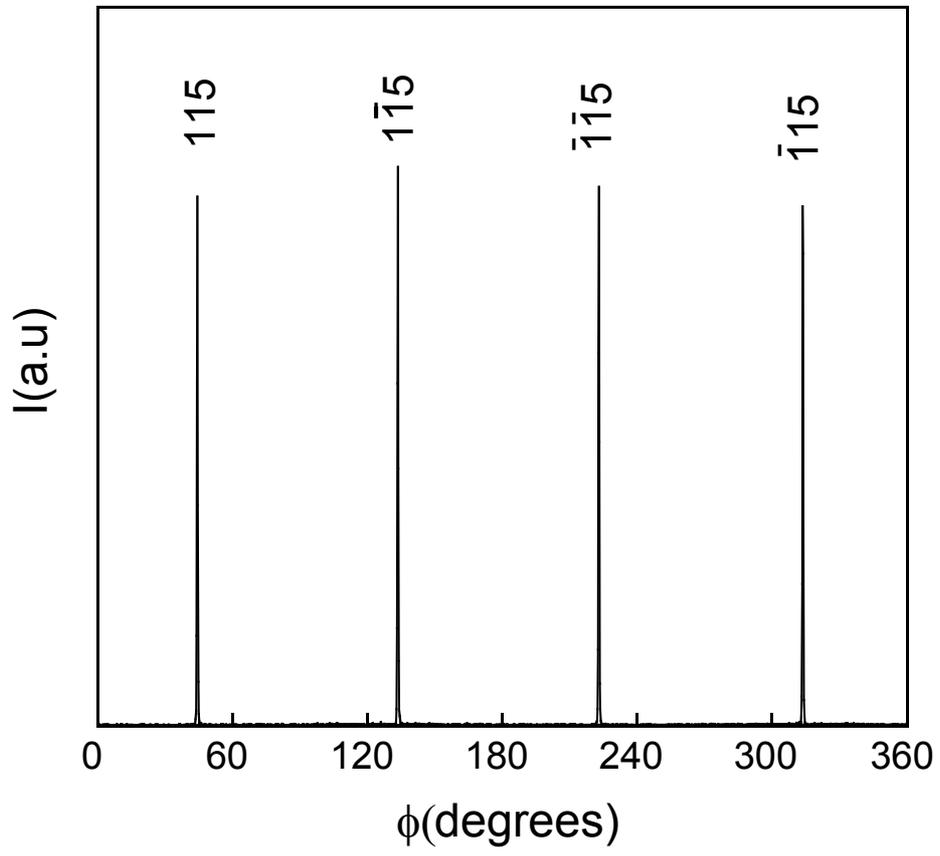

Z.Z. Li *et al.* Fig.5

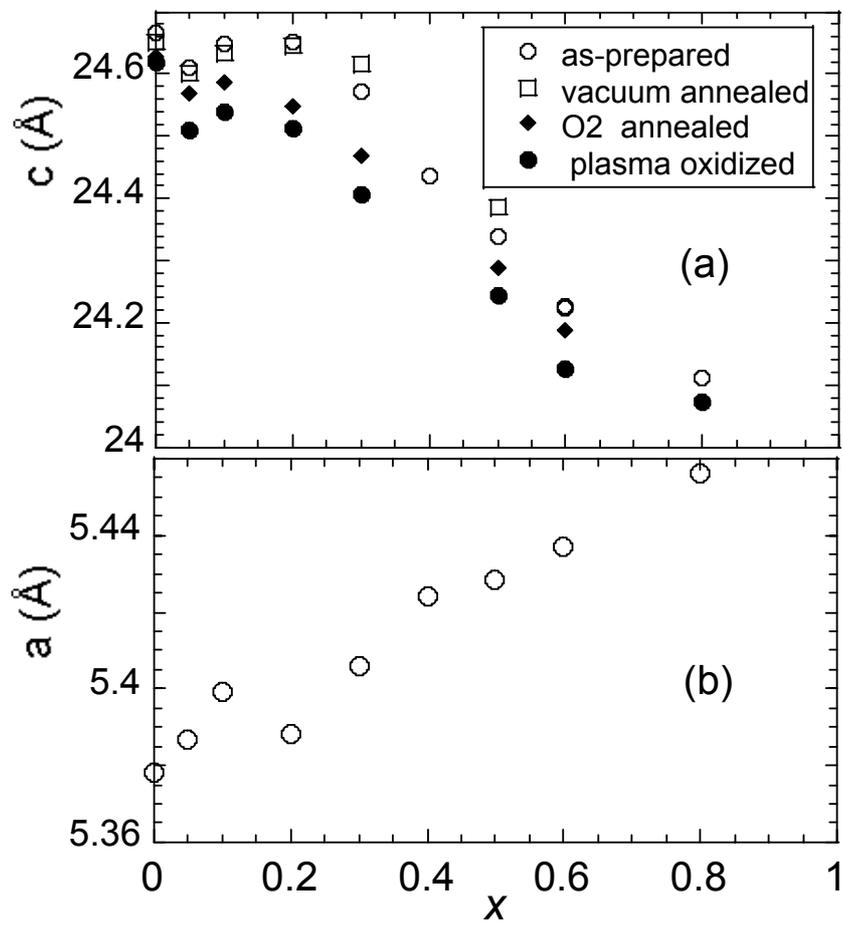

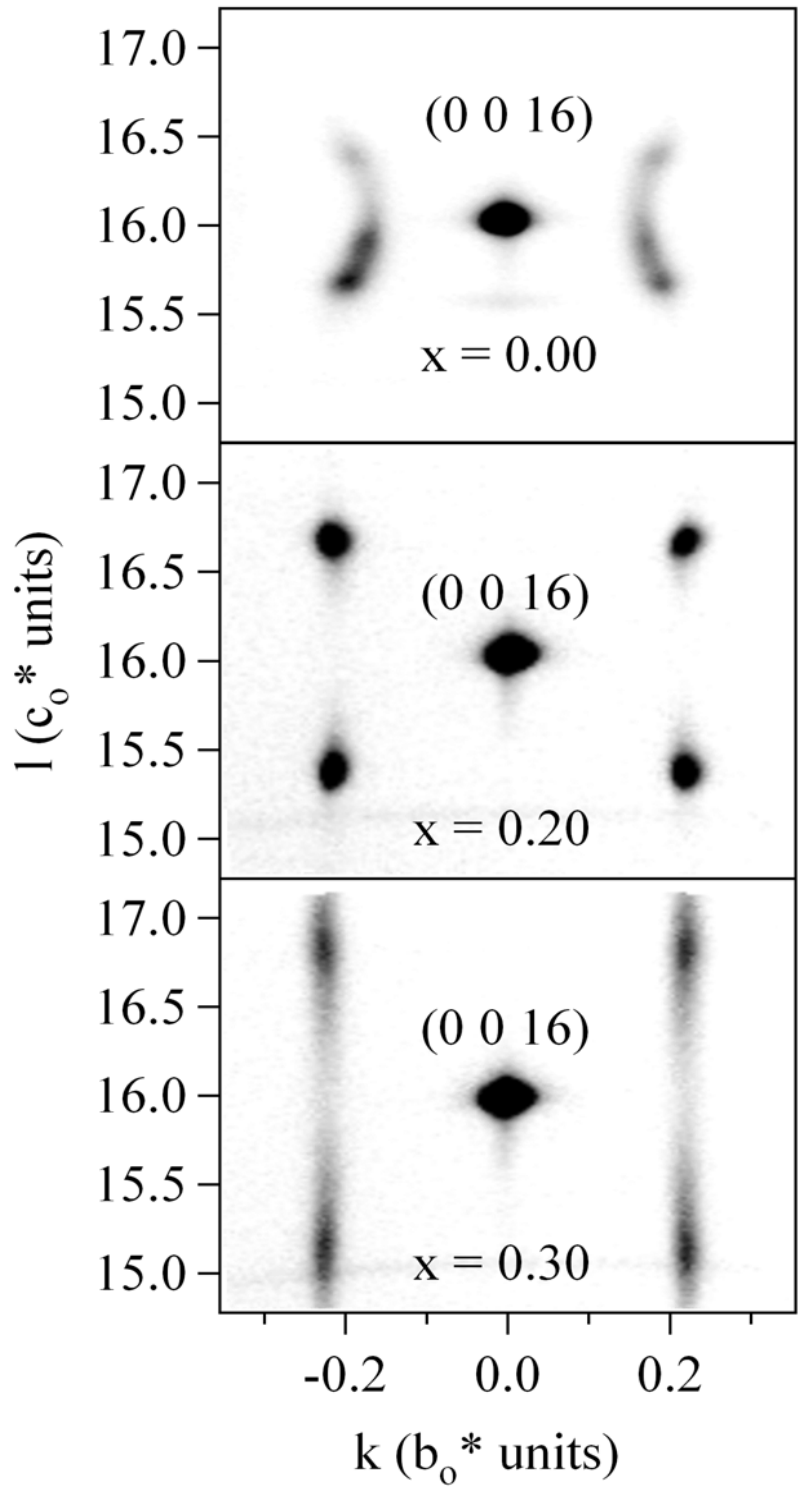

Z.Z. Li *et al.* Fig.7

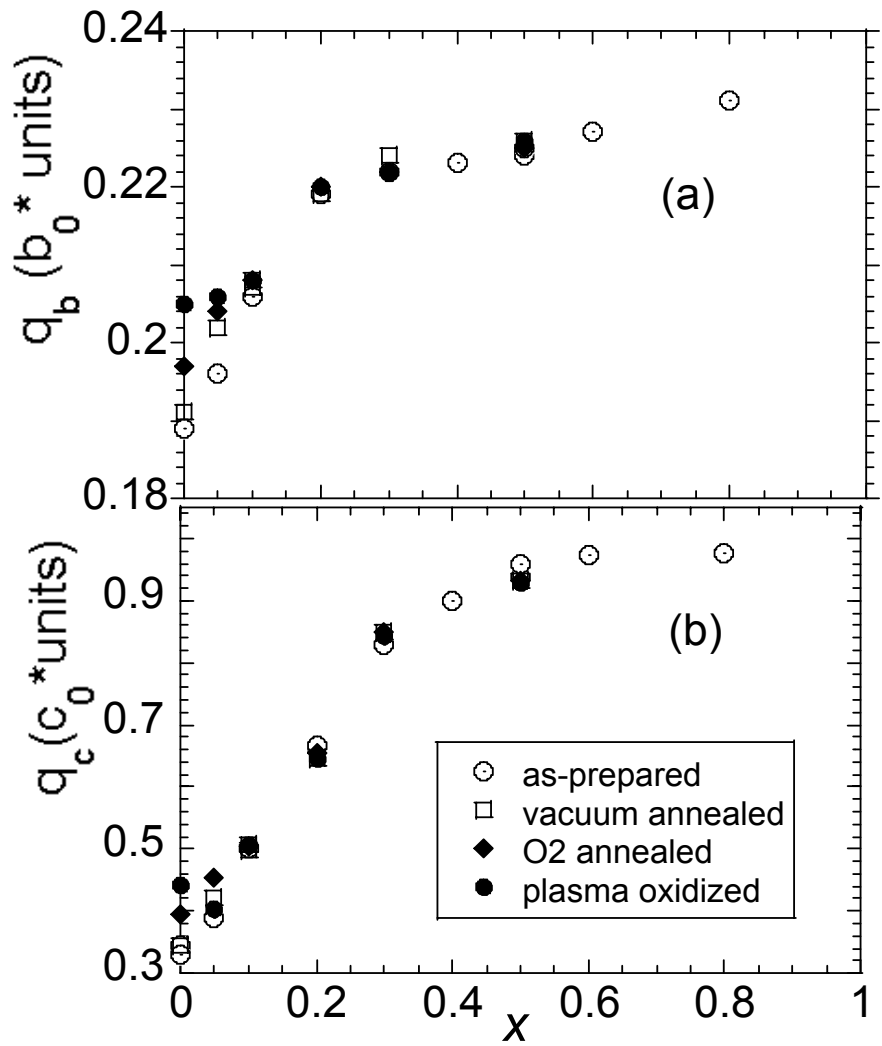

Z.Z. Li *et al.* Fig.8

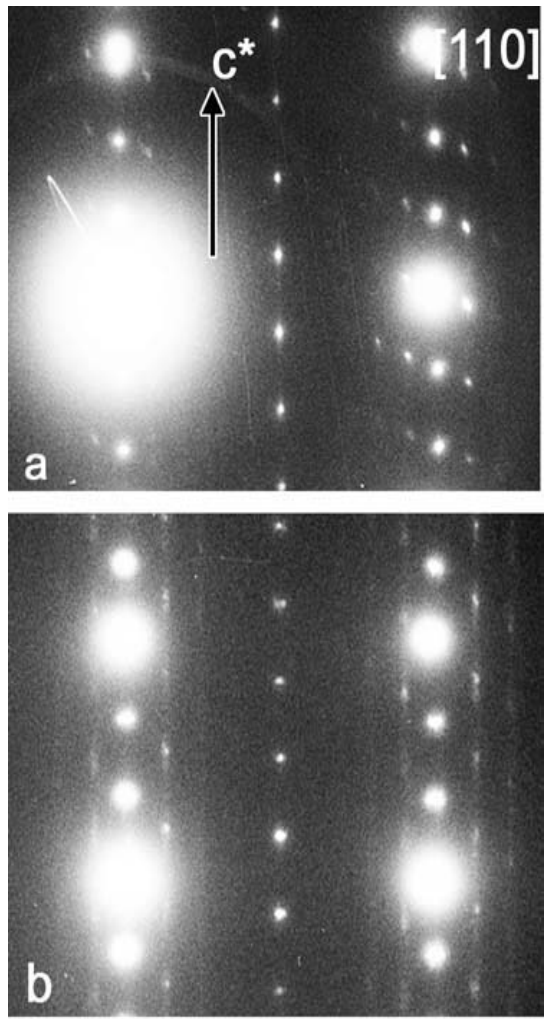

Z.Z. Li *et al.* Fig.9

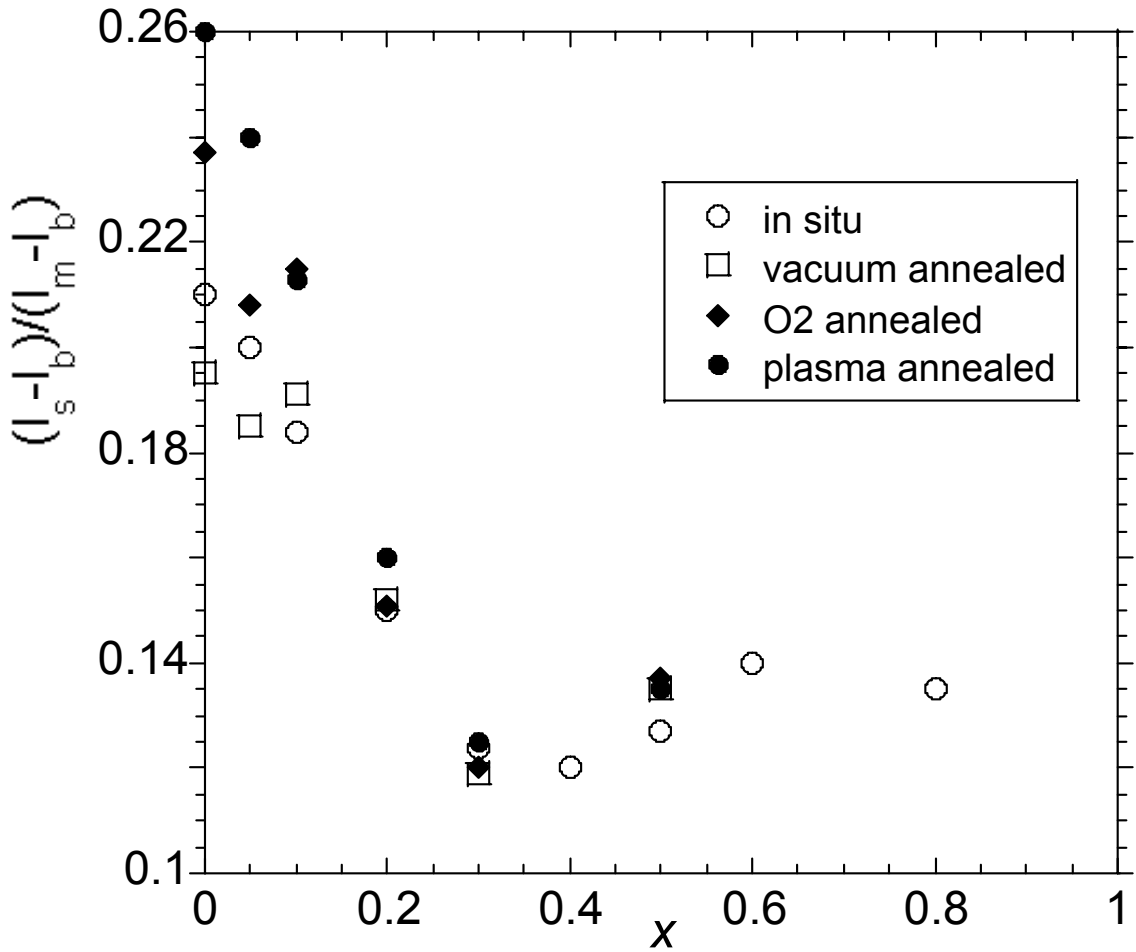

Z.Z. Li *et al.* Fig.10